\documentstyle[12pt]{article}
\def\be{\begin{equation}}
\def\ee{\end{equation}}
\def\bear{\begin{eqnarray}}
\def\eear{\end{eqnarray}}
\def\bearnn{\begin{eqnarray*}}
\def\eearnn{\end{eqnarray*}}
\def\nn{\nonumber}
\def\e8{E_8 \times E_8}
\def\spin32{Spin(32) \over {\bf Z_2}}
\def\half{1\over 2}
\def\apm{\alpha^{\prime}}
\def\Rpm{R^{\prime}}
\def\lpm{\lambda^{\prime}}
\def\s1{{\bf S}^1}
\def\BZ{{{\bf Z}}}

\def\BR{{{\bf R}}}
\def\hK{{hyper-K\"ahler}}
\newcommand\MR[1]{{{\bf R}^{#1}}}               
\newcommand\MS[1]{{{\bf S}^{#1}}}               
\newcommand\MT[1]{{{\bf T}^{#1}}}               

\newcommand\SUSY[1]{{{\cal N}= {#1}}}           
\newcommand\SLZ[1]{{SL({#1},\bf Z)}}              

\def\a{{\alpha}}

\def\g{{\gamma}}

\def\s{{\sigma}}

\def\l{{\lambda}}

\begin{document}
\begin{titlepage}
\rightline{NSF-ITP-99-135}
\begin{centering}
 
{\ }\vspace{2cm}
 
{\Large\bf S-duality and Tensionless 5-branes in Compactified 
           Heterotic String Theory\\}
\vspace{2cm}
Morten Krogh\\
\vspace{1.0cm}
{\em Institute for Theoretical Physics}\\
{\em University of California, Santa Barbara, CA 93106}\\
{\em USA}\\
{ E-mail: {\tt krogh@itp.ucsb.edu}}\\
\vspace{1cm}
\begin{abstract}
\baselineskip=16pt
\noindent
We give a simple proof of the known S-duality of Heterotic 
String theory  compactified on a $\MT{6}$. Using this 
S-duality we calculate the tensions for a class of BPS 5-branes 
in Heterotic String theory on a $\MS{1}$. One of these, the 
Kaluza-Klein monopole, becomes tensionless when the radius of the 
$\MS{1}$ is equal to the string length. We study the question of 
stability of the Heterotic NS5-brane with a transverse circle. 
For large radii the NS5-brane is absolutely stable. However 
for small radii it is only marginally stable. We also study the 
moduli space of 2 Kaluza-Klein monopoles and show that it is equal 
to the moduli space of a Heterotic $A_1$ singularity.   

\end{abstract}

\end{centering} 

\vspace{1.1cm}

\noindent November 1999

\end{titlepage}
\baselineskip=18pt
\section*{Introduction} 
In this paper we study various aspects of Heterotic string theory 
\cite{GrossRohm}. 
Heterotic string theory compactified on $\MT{6}$ has a non-perturbative 
$\SLZ{2}$ duality. The duality was first presented by Font, Ibanez, 
Lust and Quevedo in \cite{Lust}.
Evidence 
in favor of it was presented by A. Sen in \cite{SenHetsix}. 
Later the duality was proven \cite{Sen9802051} by relating it to the 
duality between Type IIA on K3 and Heterotic on $\MT{4}$ \cite{HullTown}.
Here we present a simple proof of this strong-weak coupling duality 
 using only the fact that M-theory on a circle is type IIA and 
T-dualities in type II. 

We explore the consequences of this duality for Heterotic string 
theory on a circle with unbroken gauge group. We find an exact formula 
for the tension of the Kaluza-Klein monopole. The result is 
that the tension is 
\be
T = | {m_s^6 \over \l^2} - {m_s^8 R^2 \over \l^2}| 
\ee
where $m_s$ is the string mass, $R$ is the radius of the circle 
and $\l$ is the Heterotic coupling constant. We see that at the 
selfdual radius, $m_sR=1$, this 5-brane becomes tensionless. 
This has some equivalent statements in some dual systems. 
For instance compactifying this further on a $\MT{3}$ this system is dual to 
type IIA on $K3$ with a $D6$-brane wrapped on the $K3$. The 
statement here is that the resulting 2-brane becomes tensionless 
for a certain value of the volume of the $K3$.

We also study the stability of a Heterotic NS5-brane with a transverse 
circle. It turns out that for large radii the NS5-brane is 
completely stable, but for radii smaller than the selfdual radius the 
NS5-brane is only marginally stable. It can split into two 5-branes, 
whose sum of tensions is equal to the NS5-brane tension. 

Finally we discuss the moduli space of two Kaluza-Klein 
monopoles. The moduli space of ADE singularities has recently 
been studied \cite{WitHetADE,Rozali,AspPles,Mayr}. In the case of 
an $A_1$ singularity the moduli space was found to be the Atiyah-Hitchin 
space. We show explicitly following the methods of 
\cite{WitHetADE} that the moduli space of two Kaluza-Klein 
monopoles is still the Atiyah-Hitchin space. In other words the transverse 
circle does not change the moduli space.

The organisation of the paper is as follows. In section(1) we 
explain and prove the strong-weak coupling duality of 4-dimensional 
Heterotic theory. In section(2) we study a certain class of 5-branes 
of the Heterotic theory on a circle and find the tensions for those 
5-branes which are BPS. 
We find that a certain 5-brane becomes tensionless when the radius of 
the compactification circle is selfdual under T-duality.
In section(3) we identify this 5-brane as the Taub-Nut solution, 
also called the Kaluza-Klein monopole. In section(4) we study 
the question of stability of the NS5-brane in 9-dimensional Heterotic 
theory. In section(5) we study the moduli space of two Kaluza-Klein 
monopoles.

\section{Proof of \SLZ{2} duality}
Heterotic string theory compactified on a $\MT{6}$ is known to 
have a moduli space which is 
\bear
&&(O(22,\BR ) \times O(6,\BR )) \backslash O(22,6,\BR ) / O(22,6,\BZ) 
\times \nn \\
&& SO(2,\BR ) \backslash SL(2,\BR ) / \SLZ{2} \nn
\eear
The U-duality group is 
$$
O(22,6,{\bf Z}) \times \SLZ{2}
$$
The $O(22,6,{\bf Z})$ is the perturbative T-duality group. 
The $\SLZ{2}$ is a non-perturbative duality. Evidence 
in favor of it was presented by A. Sen in \cite{SenHetsix}. 
The evidence consisted of four pieces: 
\begin{itemize}
\item
The low energy equations of motion had the symmetry.
\item
The allowed spectrum of electric and magnetic charges 
was consistent with the duality.
\item
Agreement of certain BPS masses in 4 dimensions. 
\item
Certain Yukawa couplings of BPS saturated fields 
could be checked explicitly.
\end{itemize}
Later the duality was proven \cite{Sen9802051} by relating it to the 
duality between Type IIA on K3 and Heterotic on $\MT{4}$ \cite{HullTown}.
By compactifying this duality further on a $\MT{2}$ the non perturbative 
$\SLZ{2}$ on the Heterotic side could be identified with one of the $\SLZ{2}$ 
factors in the T-duality of the $\MT{2}$ on the Type IIA side.

The aim of this section is to prove the existence of this 
$\SLZ{2}$ duality in the Heterotic theory on $\MT{6}$
in a simpler way. 

The two different Heterotic theories in 10 dimensions are 
dual when compactified on a circle and certainly also on a $\MT{6}$, 
so we do not have to distinguish between the two. 
The moduli of the compactification can 
be parametrised as follows. There is the metric on $\MT{6}$, the 
$B$-field on $\MT{6}$, $\e8$ Wilson lines on the 6 circles and the 
4 dimensional string coupling, $\lambda$. Furthermore, and of 
central importance to this $\SLZ{2}$ duality, there is a two-form
in 4 dimensions coming from the NS-NS $B$-field in 10 dimensions. 
In 4 dimensions a two-form is dual to a scalar. Hence we get another 
scalar, $a$, called the axion. Another way of understanding the scalar 
$a$ is to dualise the $B$-field in ten dimensions to a six-form 
$B^{(6)}$. The integral of this six-form on $\MT{6}$ gives a scalar in 
4 dimensions. 
We can conveniently collect the string coupling, $\lambda$, and the 
axion into a complex number in the upper halfplane,
$$
\tau = a + i {1 \over \l}
$$
The $\SLZ{2}$ duality acts as follows. It leaves the metric, the 
$B^{2}$-field on $\MT{6}$ and the Wilson lines invariant and it acts 
on $\tau$ as
$$
\tau \rightarrow {p \tau + q \over r \tau + s}
$$
where 
\be
\pmatrix{ p & q \cr r & s \cr} \in \SLZ{2} \nn
\ee
It is well known that \SLZ{2} is generated by the two elements
\be
S = \pmatrix{ 0 & -1 \cr 1 & 0 \cr } \nn
\ee
and
\be
T = \pmatrix{1 & 1 \cr 0 & 1 \cr }  \nn
\ee
To show the full $\SLZ{2}$ duality we thus just have to 
show that $S$ and $T$ are dualities. $T$ is actually still 
perturbative, since it it does not touch the coupling. $T$ 
shifts the axion. This is a symmetry as follows from the gauge 
symmetry of the six-form, $B^{(6)}$ in ten dimensions. In four 
dimensions there are instantons coming from Euclidean NS5-branes wrapped 
on $\MT{6}$. They are weighted with $e^{i2\pi a}$ in the action, showing 
that $a \rightarrow a+1$ is a symmetry. It is similar to the 
well known $2\pi$ shift of the $\theta$-angle in 4 dimensional 
gauge theories.  

The non trivial point is to establish existence of the 
strong-weak duality, $S$. To do that we will use the 
Horava-Witten picture of $\e8$ Heterotic string theory \cite{HorWit}.
By compactifying further on a $\MT{6}$ we see that the Heterotic string
 on $\MT{6}$ can be represented as M-theory on $\MT{7}$ modded out by 
a $\BZ_{2}$ symmetry. To be more precise, consider M-theory on 
$\MR{1,3} \times \MT{7}$. The coordinates on $\MT{7}$ are called 
$x^5 , x^6 , \dots , x^{11}$. M-theory has a 3-form field, $C^{(3)}$. 
The $\BZ_{2}$ group is generated by an element $\omega$ which acts as 
follows. $\omega$ is a parity operation on $x^{11}$ and changes the  
sign of $C^{(3)}$. The parity operation alone is not a symmetry of 
M-theory because of the Chern-Simons term in 11-dimensional 
supergravity. We will now perform a certain U-duality transformation, say U, 
of M-theory on $\MT{7}$. Then we will obtain M-theory on another $\MT{7}$. 
The symmetry $\omega$ will be mapped to another symmetry
$$
\tilde{\omega} = U \omega U^{-1}
$$
We now mod out by the respective symmetry on 
both sides and identify the result. 
In other words M-theory on $\MT{7}$ modded out by $\omega$ is 
dual to M-theory on the dual $\MT{7}$ modded out by $\tilde{\omega}$.

We will now define the relevant U-duality transformation and track the 
transformation of the fields. We start with M-theory on $\MT{7}$ with 
radii $R_5 , R_6 , \dots , R_{11}$. The 11-dimensional 
Planck Mass is called $M_{P}$. We will take the torus to be rectangular 
for simplicity. The chain of dualities is as follows. 
\begin{itemize}
\item{(1)}

Go to Type IIA by using the 5'th circle as 
the compactification circle.
The string mass, $m_s$, coupling, $\l$ and radii, $r_I$ are
\bear
m_s &=& (M_{P}^3 R_5)^{\half} \nn \\
\l &=& (M_{P}R_5)^{3 \over 2} \nn \\ 
r_I &=& R_I , \qquad I=6,7,8,9,10,11 \nn
\eear
\item{(2)}

Perform T-duality on the 5 circles with radii $R_6, \dots, R_{10}$ 
to get to type IIB with parameters
\bear
m_s &=& (M_{P}^3 R_5)^{\half} \nn \\
\l &=& {1 \over M_{P}^6 R_5 R_6 R_7 R_8 R_9 R_{10}} \nn \\ 
r_I &=& { 1\over M_{P}^3 R_5 R_I} , \qquad I=6,7,8,9,10 \nn \\
r_{11} &=& R_{11} \nn
\eear
\item{(3)}

Next we perform an S-duality of type IIB. The parameters become
\bear
m_s &=& M_{P}^{9 \over 2} R_5 (R_6 R_7 R_8 R_9 R_10)^{\half} \nn  \\
\l &=& M_{P}^6 R_5 R_6 R_7 R_8 R_9 R_{10} \nn \\ 
r_I &=& { 1 \over M_{P}^3 R_5 R_I} , \qquad I=6,7,8,9,10 \nn \\
r_{11} &=& R_{11} \nn
\eear
\item{(4)}

This is followed by another T-duality on the 5 circles with 
radii \linebreak $R_6, \dots, R_{10}$ which brings us back to type IIA 
with parameters
\bear
m_s &=& M_{P}^{9 \over 2} R_5 (R_6 R_7 R_8 R_9 R_10)^{\half} \nn  \\
\l &=& {R_5 \over M_{P}^{3 \over 2} (R_6 R_7 R_8 R_9 R_{10})^{\half}} \nn \\ 
r_I&=&{ R_I \over M_{P}^6 R_5 R_6 R_7 R_8 R_9 R_{10}},
 \qquad I=6,7,8,9,10 \nn \\
r_{11} &=& R_{11} \nn
\eear
\item{(5)}

Finally we lift this up to M-theory by opening up a circle which is called 
the 5'th circle again. We arrive at M-theory on $\MT{7}$ with parameters 
\bear
\tilde{M_{P}} &=& M_{P}^5 (R_5 R_6 R_7 R_8 R_9 R_{10})^{2 \over 3} \nn \\ 
\tilde{R_{I}} &=& {R_I \over M_{P}^6 R_5 R_6 R_7 R_8 R_9 R_{10}}, 
 \qquad I=5,6,7,8,9,10 \label{mrelation} \nn \\
\tilde{R_{11}} &=& R_{11} \nn
\eear
\end{itemize}
We now have two different $\MT{7}$ compactifications of M-theory which 
are dual. Actually the eleventh circle was not touched at all in this 
chain of dualities, so the duality is already a duality of M-theory 
on $\MT{6}$. We now want to mod out by respectively $\omega$ and 
$\tilde{\omega}$ and identify the resulting theories. To do that 
we need to find $\tilde{\omega}$. In order to do  that we must track 
the various fields of M-theory under the duality. The bosonic fields 
are the metric $g_{IJ}$ and the 3-form $C_{IJK}$ and the 
dual 6-form $C_{IJKLMN}$. The 3-form and 6-form  are related 
but it is still an advantage 
to keep track of both as we will see below. 
Under $\omega$ they transform as 
\begin{eqnarray*}
g_{IJ} & \rightarrow & g_{IJ} \qquad \;\; I,J=1,\dots,10 \\
g_{I(11)}& \rightarrow & - g_{I(11)} \qquad \; I=1,\dots, 10  \\
C_{IJK}& \rightarrow &-C_{IJK} \qquad \; I,J,K=1,\dots,10 \\
C_{IJ(11)}& \rightarrow & C_{IJ(11)} \qquad \; I,J =1,\dots, 10  \\
C_{IJKLMN}& \rightarrow &C_{IJKLMN} \qquad I,J,K,L,M,N=1,\dots,10 \\
C_{IJKLM(11)}& \rightarrow & - C_{IJKLM(11)} \qquad I,J,K,L,M =1,\dots, 10  
\end{eqnarray*}
The transformation law of the 6-form follows from the relation
\be
dC^{(3)} = * dC^{(6)} \nn
\ee
We now claim that 
under the U-duality transformation a field that transforms 
with eigenvalue 1 is mapped into another field
with eigenvalue 1 and the same is true for eigenvalue -1. 
To see that one has to track all the different fields through 
the chain of dualities. Let us here just do it explicitly for 
one case, namely $g_{1(10)}$. In the first step above it 
remains $g_{1(10)}$. In the second step it becomes $B^{NS}_{1(10)}$. 
In the third step it becomes $B^{RR}_{1(10)}$. In the 4th step 
it becomes $C^{RR}_{16789}$. In the 5th step it becomes 
$C_{156789}$. It is indeed true that both  $g_{1(10)}$ and
 $C_{156789}$ have the same eigenvalue, namely 1.

This implies that $\tilde{\omega} = \omega$. We conclude that 
by modding out on both sides by $\omega$ we get a 
duality between two Heterotic theories on $\MT{6}$. 
In the Horava-Witten picture the relation between the M-theory 
parameters, $M_{P}$ and the radius of the eleventh circle $R_{11}$, 
and the Heterotic parameters, $m_s$ and $\l$ are
\bearnn
m_s^2 &=& M_P^3 R \\
\l &=& (M_P R)^{3 \over 2} 
\eearnn
Using these relations and Eq.(\ref{mrelation}) 
we can relate the parameters of the 
Heterotic theory on 
one side of the duality with those on the other side. 
\bear
\tilde{m_s} &=& { m_s^7 R_5 R_6 R_7 R_8 R_9 R_{10} \over \l^2} \nn \\
\tilde{\l}  &=& { m_s^6  R_5 R_6 R_7 R_8 R_9 R_{10} \over \l}\label{hetrel}\\
\tilde{R_I} &=& { \l^2 R_I \over  m_s^6  R_5 R_6 R_7 R_8 R_9 R_{10}}
\qquad I=5,6,7,8,9,10  \nn
\eear
This looks a bit messy but we can understand it better by 
expressing it in the right variables. In the moduli space 
\bear
&&(O(22,\BR ) \times O(6,\BR )) \backslash O(22,6,\BR ) / O(22,6,\BZ) 
\times \nn \\
&& SO(2,\BR ) \backslash SL(2,\BR ) / \SLZ{2} \nn
\eear
the radii of the torus are contained in the first factor
and they are expressed in string units. We easily calculate 
that 
\be
\tilde{m_2} \tilde{R_I} = m_s R_I \qquad I=5,6,7,8,9,10 \nn
\ee
This shows that the radii are unchanged in the right units. 
We could have worked with a non-rectangular torus if we wanted to. 
Since the rectangular torus is unchanged it is clear that any torus 
would be left unchanged. 

The second factor of the moduli space is a complex number in the 
upper halfplane, $\tau$. The imaginary part of $\tau$ is $1 \over 
\l_4$ where $\l_4$ is the four-dimensional string coupling. 
The relation between the four-dimensional and ten dimensional 
coupling is
\be
{1 \over \l_4^2 } = {m_s^6 R_5R_6R_7R_8R_9R_{10} \over \l^2} \nn
\ee
Using Eq.(\ref{hetrel}) we find that 
\be
\tilde{\l_4} = { 1 \over \l_4}
\ee
This derivation was done with zero axion, and hence is exactly 
the element, $S$, of $\SLZ{2}$. We have now derived the existence 
of this duality only for a subset of the whole moduli space. 
However the existence of a duality in just one point implies that 
it extends to the whole moduli space. Its action is to leave 
the point in the first factor invariant and act as 
\be
\tau \rightarrow {- 1 \over \tau} 
\ee
on the second factor. In particular the Wilson lines of $\e8$ are 
left invariant. In this derivation the Wilson lines could not be 
tracked since they only exist after modding out. This finishes our 
proof of the strong-weak duality. In the next section we will study 
some consequences of it.  

\section{Tensionless 5-branes in Heterotic string 
         theory on a circle}

In this section we will study the Heterotic theory on a circle 
of radius $R$. 
The gauge group could be either $\e8$ or $\spin32$. For 
simplicity we take the Wilosn line on the circle to be zero 
so the gauge group is unbroken. There are two $U(1)$ gauge fields 
in 9 dimensions, namely $g_{\mu 10}$ and $B_{\mu 10}$. 
The charges under these are respectively momentum amd winding. 
In perturbative string theory one can calculate the mass of ligthest 
particle with momentum $m$ and winding $n$. However it is only 
for certain $m$ and $n$ that the state is BPS. In the BPS case the 
mass calculated at string tree-level is exact. The quantum numbers 
and masses in the BPS cases are \cite{Ginsparg}:
\begin{itemize}
\item{(1)}
$$
m \geq 0, n \leq 0, \;\;\; {\rm or }\;\;\;  m \leq 0, n \geq 0,  
\qquad M= |{m \over R} - n m_s^2 R|   
$$
\item{(2)}
$$
m=1 , n=1, \;\;\; {\rm or} \;\;\;  m=-1 , n=-1, 
 \qquad M = |{1 \over R} - m_s^2 R |    
$$
\end{itemize}
These conditions secure that there are no excited rightmoving 
oscillators. The last two states become massless at the selfdual 
radius. They are the W-bosons of the enhanced $SU(2)$ gauge symmetry. 
By compactifying further on a $\MT{5}$ these particles still exist 
with the same masses. We can now apply the S-duality transformation 
discussed in the previous section. Let the radii of the $\MT{5}$ be 
$R_5,R_6,R_7,R_8,R_9$. The S-dual object of the particle with 
quantum numbers $(m,n)$ is a particle with a mass which is easily 
calculated to be.
\be
M = | m {m_s^6 \over \l^2} - n {m_s^8 R^2 \over \l^2}| 
    R_5 R_6 R_7 R_8 R_9 
\ee
We see that it is a particle with a mass proportional to the 
volume of the $\MT{5}$. This implies that it is a 5-brane in 
9 dimensions with a tension of  
\be
T_{(m,n)} =  | m {m_s^6 \over \l^2} - n {m_s^8 R^2 \over \l^2}| 
\ee
These 5-branes are magnetically charged under the two 
$U(1)$ gauge fields in 9 dimensions. The quantum number 
$m$ denotes NS5-brane charge. An NS5-brane indeed has tension 
\be
T_{(1,0)} = {m_s^6 \over \l^2} 
\ee
The quantum number $n$ denotes the Kaluza-Klein monopole charge. 
In other words the circle of the compactification can be in a 
non-trivial bundle over the transverse $\MR{3}$ at infinity. 
$n$ measures the first chern class of this bundle. 
The 5-branes are BPS whenever the dual particles are. 
Let us look closer at one of the special cases listed above,$m=1,n=1$. 
The other one $m=-1,n=-1$ is just the antibrane. 
This BPS 5-brane has a tension which is 
\be
T_{(1,1)} = | {m_s^6 \over \l^2} - {m_s^8 R^2 \over \l^2}| 
\ee
We see that for large $R$ it is heavy but for the selfdual radius 
$m_s R =1$ it is tensionless. What is this 5-brane? In the next 
section we will see that it is the Taub-Nut space solution. 

Let us see how the two kinds of branes 
transform under T-duality.
Heterotic string theory with gauge group, G, on a circle 
without a Wilson line has a T-duality that preserves the gauge 
group. The relation between the original parameters and the new ones 
is
\bear
\Rpm &=& {1\over m_s^2 R} \label{tparam} \\
\lpm &=& {\lambda \over m_s R} \nn
\eear
with the same $m_s$. There is a gauge field, $g_{\mu 10}$, in 9 dimensions 
coming from the metric and a gauge field, $B_{\mu 10}$, coming from the 
$B$-field. The electric charge for  $g_{\mu 10}$ is momentum 
around the circle. 
The magnetic charge is Kaluza-Klein monopole charge. The electric charge 
for $B_{\mu 10}$ is string winding 
and the magnetic charge is NS5-brane charge.
These two gauge fields are swapped under the T-duality. Momentum and 
winding are swapped and Kaluza-Klein and NS5-brane charges are swapped.
Under T-duality of the $\MS{1}$ the quantum numbers $m,n$ are mapped 
as follows
$$
m \rightarrow -n , \;\;\;\;  n \rightarrow -m
$$
Especially the Taub-Nut space is mapped to the Taub-Nut space 
with opposite orientation.
$$
(m,n)=(1,1) \rightarrow (m,n)=(-1,-1) 
$$

\section{The Taub-Nut solution in the Heterotic  
         String theory on a circle}
In this section we will discuss the supergravity solution for 
the Taub-Nut 5-brane in the Heterotic string 
theory compactified 
on a circle down to 9 dimensions. Supergravity solutions 
for monopoles in compactified Heterotic string theories have been discussed 
extensively in the litterature \cite{BDDF, HarLiu, Khuri, Gaunt}. 
We review parts of the construction 
in order to understand that the brane with charges $(1,1)$ is the 
Taub-Nut 5-brane.   
The gauge group could be either $\e8$ or $\spin32$. It will 
not make a difference for our purpose. We will take no Wilson line 
on the circle and hence keep the gauge group unbroken. 
By 5-brane we understand an object with 
5+1 dimensional Lorentz invariance and localised in the 3 transverse 
directions. There are two basic types of these, namely the NS5-brane and 
the Kaluza-Klein monopole. Here we only discuss the Kaluza-Klein 
monopole also called Taub-Nut space.
 The NS5-brane exists already in 10 dimensions and the 
9 dimensional one is trivially obtained from the 10 dimensional one.
5-branes in 10 dimensions were discussed in \cite{CHS}.
The Kaluza-Klein monopole, on the other hand, only exists after 
compactification. Now we will discuss the supergravity solution 
of this object closely following the discussion of \cite{CHS}. 
As we will see below the Kaluza-Klein monopole preserves
half of the supersymmetry, i.e. 8 supercharges.

The field content of the Heterotic string theory in 10 dimensions 
is the metric, $\g_{MN}$, the dilaton, $\phi$, a two-form 
poten\-tial, $B$, a gauge potential, $A$, a gaugino $\chi$, 
 a dilatino, $\lambda$ and a gravitino, $\psi_M$. 
The bosonic part of the action in 10 dimensions reads 
\be
S = {1 \over 2 \kappa^2}\int d^{10}x \sqrt{-g} e^{-2 \phi} \left(
                R + 4(\nabla \phi)^2 - {1 \over 3} H^2 -
                {\apm \over 30} {\rm Tr} F^2 \right) 
\ee
where the three-form antisymmetric tensor field strength is related to the
 two-form potential by the familiar anomaly equation \cite{gsanom}.
\be
H=dB +\apm\left(\omega^L_3(\Omega)-{1\over 30}
			\omega^{YM}_3(A)\right) \label{Hligning}
\ee
where $\omega_3$ is the Chern-Simons three-form. This equation 
implies
\be
dH=\apm (trR \wedge R- {1 \over 30}Tr F \wedge F)   \label{bianchi}
\ee
The trace $Tr$ is taken in the adjoint representation of the gauge group.
The connection
$\Omega$ appearing in Eq.(\ref{Hligning}) is a 
non-Riemannian connection related
to the usual spin connection $\omega$ by
\be
\Omega_{M}^{AB} = \omega_M^{~AB} -  H_M^{~AB}
\ee
In these equations $M,N$ are Einstein indices and $A,B$ are Lorentz 
indices.
The gravitational constant,$\kappa$, is related to $\apm$ and the 
string coupling, $\lambda$, by 
\be
{1 \over 2\kappa^2} = {1 \over (2\pi)^7 {\apm}^4 \lambda^2}
\ee
We will be interested in solutions that preserve some supersymmetry. 
In other words the supersymmetry variation of the fermions 
should be zero. 
The fermion
supersymmetry transformation laws are
\bear
\delta \chi &=& F_{MN} \Gamma^{MN} \epsilon \\
\delta \lambda &=& (\Gamma^{M} \partial_M \phi -
         {1 \over 6} H_{MNP}\Gamma^{MNP}) \epsilon \label{susytrans}  \\
\delta \psi_M &=& (\partial_M + {1 \over 4}\Omega_{-M}^{AB}
                   \Gamma_{AB}) \epsilon   
\eear
Here $\epsilon$ is the infinitesimal supersymmetry parameter. 
$\epsilon$ is a Majorana-Weyl spinor of $SO(1,9)$. It has 
16 real components. 

Let the radius of $\MS{1}$ be $R$.
The metric is a product of flat ${\bf R}^{1,5}$ and the 4 dimensional 
Taub-Nut metric
\be
ds^2 = R^2 U(dy - A_i dx^i )^2 + U^{-1}(d\vec{x})^2,
\qquad
i=1,2,3,\qquad 0\le y\le 2\pi
\label{tnmet}
\ee
where,
$$
U = \left(1 + {R \over {2|\vec{x}|}}\right)^{-1},
$$
and $A_i$ is the gauge field of a monopole centered at the origin.
The Taub-NUT space has the following desirable properties,
\begin{itemize}
\item
If we excise the origin, what remains is a circle fibration
over ${\bf R}^3 -\{0\}$. Eq.(\ref{tnmet}) is written such that
$\vec{x}$ is the coordinate on this base ${\bf R}^3 -\{0\}$.
 For $|\vec{x}|$ restricted to a constant, the fibration is
exactly the Hopf fibration of $\MS{3}$ over $\MS{2}$.
\item
The origin $\vec{x} = 0$ is a smooth point.
\item
As $|\vec{x}|\rightarrow\infty$ the radius of the fiber
becomes $R$.
\item
The space has a $U(1)$ isometry group that preserves the
origin $\vec{x}=0$. An element $g(\theta) = e^{i\theta}\in U(1)$
acts by $y\rightarrow y+\theta$.
\end{itemize}
We will now show that this space 
 with a constant dilaton and 
 all other fields set to zero is a supersymmetry preserving 
solution to zero'th order in $\apm$. Firstly Eq.(\ref{Hligning}) 
is trivially satisfied to zero'th order in $\apm$. 
The supersymmetry variation of $\chi$ and $\lambda$ are zero 
 since the right hand sides are zero for any $\epsilon$.
The only nontrivial equation is the gravitino variation which 
now reads 
\be
\delta \psi_M = (\partial_M + {1 \over 4}\omega_M^{~AB}\Gamma_{AB}) 
 \epsilon =  {\nabla}_M \epsilon
\ee
where $\nabla$ is the covariant derivative in the Taub-Nut metric. 
Taub-Nut space is a noncompact Calabi-Yau manifold. It has $SU(2)$ 
holonomy, which implies that there are covariantly constant spinors of 
one chirality, say positive. More concretely the spin connection 
satisfies 
\be
\omega_M^{~AB} = \epsilon^{ABCD} \omega_M^{~CD}
\ee
where $\epsilon^{ABCD}$ is the totally antisymmetric symbol with 
$\epsilon^{1234}=1$. Positive chirality means that 
\be 
\Gamma_{1234} \epsilon_+ = \epsilon_+
\ee
Combining the three equations above we get
\be
\delta \psi_M = \partial_M \epsilon_+
\ee
This is zero for a constant $\epsilon_+$. There are no unbroken 
supersymmetries of negative chirality. 
Since the original supersymmetry was a Majorana-Weyl spinor 
of positive chirality in 9+1 dimensions the unbroken supersymmetry 
is a positive chirality spinor in the 5+1 dimensions along the 
worldvolume of the brane. To higher order in $\apm$ the solution 
changes. For instance Eq.(\ref{Hligning}) implies that $H$ is 
nonzero to first order in $\apm$. The variation of the dilatino,
$\delta \lambda$, then implies that the dilaton becomes non constant 
to first order in $\apm$. One should imagine correcting the solution 
order by order in $\apm$. To do that consistently we should also 
include higer order corrections in the original action. The point is 
however that there is a solution to all orders in $\apm$ and the 
string coupling too for that sake. The exact solution will preserve 
supersymmetry. This is so because a BPS state comes in a small 
representation of the supersymmetry algebra. The size of a representation 
cannot jump when the parameters of the theory are changed continously. 
Until we have a full non perturbative formulation of string theory 
the only way to specify solutions is to do it in some approximation, 
like weak coupling. These solutions will then give rise to exact
solutions. 

Let us find the charges of this solution. The charges are integers 
and can not jump with $\apm$ so we should be able to identify the 
charges working to lowest non-trivial order. As we saw above there 
are two relevant 5-brane charges,$m$ and $n$. $m$ denotes the 
magnetic charge of the 3-form field, 
\be
m = {1 \over 16 {\pi}^2 \apm} \int_{\MS{2} \times \MS{1}} H 
\ee
$n$ denotes the first chern class of the circle-fibration at infinity. 
In this case $n=1$. What about $m$. In type II theories $m$ would be 
zero but it is different in the Heterotic theory due to the Bianchi 
identity Eq.(\ref{bianchi}). The Bianchi identity implies that the 
curvature induces magnetic charge of the $B$-field. We get 
\be
m = {1 \over 16 {\pi}^2} \int_{\MS{2} \times \MS{1}} 
    \omega_3^L(\Omega) = 1
\ee
The higher order $\apm$ corrections cannot change an integer as 
already noted. We thus conclude that the Taub-Nut space is the 
5-brane discussed in the previous section with quantum numbers
$(1,1)$. Since it preserves some supersymmetry
 its tension is given exactly by the 
formula derived in the previous section.
\be
T_{(1,1)} = | {m_s^6 \over \l^2} - {m_s^8 R^2 \over \l^2}| 
\ee
Here $m_s^2 = {1 \over \apm}$. We see that the first term 
is suppressed by $\apm$ compared to the second one. The 
second term is the ADM mass of the pure Taub-Nut solution. 
The first term would appear when taking the $\apm$ corrections 
into account.

\section{The marginally stable NS5-brane}

In this section we will study the stability of the NS5-brane
 in the Heterotic theory on a circle. 
The NS5-brane is a BPS 5-brane. It has quantum numbers 
$(m=1,n=0)$ in the notation from section 2.
It has tension
\be
T_{(1,0)} = {m_s^6 \over \l^2} 
\ee
Since it is BPS it cannot decay into 5-branes of a total tension 
which is smaller than $T_{(1,0)}$. It could be however that it could 
split into two or more 5-branes whose tension add up to $T_{(1,0)}$.
The central charge for a 5-brane with charges $(m,n)$ is 
\be
Z_{(m,n)} = m {m_s^6 \over \l^2} - n {m_s^8 R^2 \over \l^2}
\ee
The BPS inequality reads
\be
T_{(m,n)} \geq | Z_{(m,n)} |
\ee
Suppose that the NS5-brane can split into two 5-branes of 
charges $(m_1,n_1)$ and $(m_2,n_2)$. Then they must both be BPS and 
have positive central charge. 
\bearnn 
 m_1 {m_s^6 \over \l^2} - n_1 {m_s^8 R^2 \over \l^2} &\geq& 0 \\
 m_2 {m_s^6 \over \l^2} - n_2 {m_s^8 R^2 \over \l^2} &\geq& 0 \\
 m_1 + m_2 &=& 1 \\
 n_1 + n_2 &=& 0 
\eearnn
The condition that they are BPS is written down in section 2. 
It is that either $m$ and $n$ have opposite sign or they are 
both $1$ or both $-1$. It is easily seen that all these requirements 
are only satisfied if 
\bearnn
(m_1,n_1)&=& (1,1) \\
(m_2,n_2)&=& (0,-1) 
\eearnn
or vice versa and furthermore 
$$
m_s R \leq 1
$$
We thus conclude that for large radius the NS5-brane is absolutely 
stable as one would expect. However for a radius smaller than the 
selfdual radius the NS5-brane is only marginally bound. It can 
split into the 5-branes with charges $(1,1)$ and $(0,-1)$. 
The one with charge $(1,1)$ is the Taub-Nut space discussed 
in the previous section. The $(0,-1)$ 5-brane can be thought of 
as an anti Taub-Nut space bound with a NS5-brane.  

\section{The moduli space of Taub-Nut 5-branes}

In this section we will discuss the moduli space of 
several Taub-Nut 5-branes, or Kaluza-Klein monopoles, 
 in the Heterotic theory on a circle.
 In section(3) we presented the 
supergravity solution for one Taub-Nut 5-brane. 
The case of $q$ parallel Taub-Nut 5-branes is similar. The metric 
is to lowest order the $q$-centered Taub-Nut space with metric 
\be
ds^2 = R^2 U(dy - A_i dx^i )^2 + U^{-1}(d\vec{x})^2,
\qquad
i=1\dots 3,\qquad 0\le y\le 2\pi.
\label{qtnmet}
\ee
where,
$$
U = \left(1 + \sum_{l=1}^{q} {R \over {2|\vec{x}- \vec{x_l}|}}\right)^{-1},
$$
and $A_i$ is the gauge field of $q$ monopoles centered at $x_l$,
$l=1,\dots, q$. This space is smooth except when two or more centers 
coincide. For $n$ coinciding centers the space develops an ${\bf R}^4/
{\bf Z}^n$ orbifold singularity. An ${\bf R}^4/
{\bf Z}^n$ orbifold singularity is also sometimes denoted a $A_{n-1}$ 
singularity. The total charge of this system is $(q,q)$. 
A NS5-brane, which has charge $(1,0)$, can be thought of as a 
small instanton \cite{Witsmallinst}. In other words a NS5-brane 
can spread out and become an instanton in the transverse space. 
The 5-branes with charge $(q,q)$ cannot spread out. Their 
moduli space is $4q$ dimensional. $3q$ of these dimensions come from 
specifying the $q$ centers of the $q$-centered Taub-Nut space. 
The last $q$ parameters come from the $B$-field. The $q$-centered 
Taub-Nut space has $q$ normalisable harmonic two-forms. The 
$q$ scalars come from expanding the $B$-field into these $q$ 
normalisable harmonic two-forms.
 These $q$ parameters 
are periodic due to gauge symmetry of the $B$-field. It is 
equivalent of periodicity of Wilson lines of a gauge field.
$4$ of these $4q$ dimensions parametrise the center of mass motion, 
which is $\MR{3} \times \MS{1}$. 
It decouples from the rest of the moduli space. 

The moduli space of $q$ Kaluza-Klein monopoles has been argued by 
Sen to be equal to the moduli space of $q$ BPS monopoles in 
four-dimensional $\SUSY{4}$ $SU(2)$ gauge theory \cite{SenKK}. 
For coinciding 5-branes the space has an $A_{q-1}$ singularity. 
The moduli space of the Heterotic theory on a $A_{q-1}$ singularity 
has recently been studied by Witten \cite{WitHetADE}. He conjectured that it 
is equal to the moduli space of 3-dimensional $\SUSY{4}$ pure
gauge theory with gauge group $SU(q)$. Similar statements were 
conjectured for $D$ and $E$ groups. In the case of $q=2$ it was shown 
explicitly that the moduli space is equal to the Atiyah-Hitchin space. 
Shortly after, Witten's conjecture was proven by means of string dualities 
\cite{Rozali,AspPles,Mayr}. 

The purpose of this section is to prove that the moduli-space of 
the two-centered Taub-Nut space is equal to the Atiyah-Hitchin space. 
In other words the moduli space does not depend on the radius $R$ of 
the circle at infinity. This result follows from Sen's identification 
of this moduli space with a monopole moduli space, but here we will 
prove it with a different argument. 
 We will show this by following Witten's 
approach \cite{WitHetADE} and see that the argument used in the case 
of an $A_1$ singularity also goes through in the case of
two-centered Taub-Nut space. 

Firstly the moduli space is a \hK\ manifold. This is because 
it is the moduli space in a 6-dimensionsonal theory with 8 supercharges 
without gravity. There is no gravity because 
 the transverse space is noncompact the 6-dimensional gravitational 
constant is zero. Hence the moduli space is that of a globally 
supersymmetric theory, i.e. \hK. 

The moduli space is independent of the string coupling constant, 
since it is a hypermultiplet moduli space. By further compactifying to 
4 dimensions the string coupling sits in a vector multiplet moduli space 
and hence does not correct the hypermultiplet moduli space \cite{deWit}. 
It can hence be studied in Heterotic conformal theory. 

Discarding the center of mass motion the moduli space is 4 dimensional. 
When the 2 5-branes are far away from each other the moduli space 
is parametrised by the separation, $\vec{x}$, and the integral of the 
$B$-field along the exceptional divisor. To lowest order in $\apm$ 
the moduli space is just $(\MR{3} \times \MS{1})/{\bf Z}_2$. 
The ${\bf Z}_2$ is due to the indistinguishability of the 2 branes. 
Interchanging them changes the orientation on the exceptional divisor 
and hence changes the sign of the integral of the $B$-field. 

In the case of an $A_1$ singularity Witten showed that $\apm$ corrections 
change the topological structure of the moduli space at infinity. 
The circle is not in a product with $\MR{3}$ at infinity but is in 
a nontrivial circle bundle of first Chern class $-4$. Exactly the same 
is true here as we will now argue. The actual Heterotic spacetime 
is of course different in this case compared to the case of a pure 
singularity but the circle in the moduli space behaves similarly. 

When the two 5-branes are on top of each other, the Taub-Nut space has 
an exact $U(1) \times SU(2)$ symmetry. The $U(1)$ tranalates along the 
circle in Taub-Nut space. The $SU(2)$ rotates the $\MR{3}$. 
When the two 5-branes are separated the $SU(2)$ acts on their 
separation vector. So we have an exact $SU(2)$ that acts on the  
moduli space. The $U(1)$ does not act on the moduli space. 
In the case of the $A_1$ singularity there was a $Spin(4)= 
SU(2)_L \times SU(2)_R$ which acted on the space. The $SU(2)_R$ 
acts on the moduli space whereas the $SU(2)_L$ does not. 
The important point is that in our case we do still have a 
$SU(2)$ that acts on the moduli space. At infinity the moduli space 
is approximately  $(\MR{3} \times \MS{1})/{\bf Z}_2$. The $SU(2)$ 
rotates the $\MR{3}$. The question of what bundle the circle is in over 
the $\MR{3}$ at infinity is the same as asking how the $SU(2)$ acts on 
the circle. Let us fix a specific point, $\vec{x}$, in $\MR{3}$. There is a
$U(1)$ subgroup of $SU(2)$ which leaves this point invariant. This 
$U(1)$ will act on the $\MS{1}$ with a charge $k$. This $k$ is exactly 
the first chern class of the bundle. Hence we have to determine $k$. 
$k$ can be determined in the same way as in the case of the pure 
$A_1$ singularity \cite{WitHetADE}. 
Here we will consider a worldsheet instanton, 
i.e. a fundametal Euclidean string wrapped on the exceptional divisor. 
The partition function of this string should be invariant under 
the $U(1)$ which leaves $\vec{x}$ invariant. 
There are two potential non-invariances 
of this partition function. One comes from the path integral over 
worldsheet fermions which gives a factor of the Pfaffian of the 
worldsheet Dirac operator, $Pf({\cal D})$. The other is the coupling 
of the $B$-field to the fundamental string $exp(i \int B)$. 
The product of the two 
\be
 Pf({\cal D})exp(i \int B)
\ee
should be invariant under $U(1)$. Performing a $U(1)$ transformation 
$e^{i\a}$ the factor $exp(i \int B)$ get multiplied by $e^{ik \a}$, 
where $k$ is the first Chern class which we want to find. $Pf({\cal D})$ 
gets multiplied with $e^{i n \a}$ where $n$ is the total $U(1)$ charge of the 
fermionic zero modes of the Dirac operator.
 For this to be invariant we must have $k=-n$. 
Hence we just have to count the total charge of the fermionic zero modes. 
In the case of the pure $A_1$ Witten counted this to be 4. The point is 
now that the presence of the circle cannot change this since it is an integer. 
More precisely, let $R >> |\vec{x}| >> \sqrt{\apm}$. Then the situation 
is almost indistinguishable to the case of a pure singularity and $n=4$. 
Now deform $R$ continously to any value. $n$ stays an integer and hence 
$n=4$. 

Now we have established that the moduli space at infinity 
is a circle fibration over $\MR{3}$ with first chern class $-4$ and 
there is a $SU(2)$ symmetry that acts on the moduli space with 
generic three-dimensional orbits. This space should be modded out 
by ${\bf Z}_2$. The moduli space is also 
 smooth by the argument in \cite{WitHetADE}. There is no 
$U(1)$ that acts on the fiber at infinity. It is broken by 
worldsheet instanton effects. As noted above the moduli space 
is \hK . The point is now that there is a unique 
space with these properties, namely 
the Atiyah-Hitchin space \cite{AtiyahHitchin}. We thus conclude that 
the moduli space of 2 Kaluza-Klein 5-branes is the Atiyah-Hitchin space.

\section*{Acknowledgements}
This work was supported by the National Science Foundation grant 
PHY 94-07194.

\clearpage

\def\np#1#2#3{{\it Nucl.\ Phys.} {\bf B#1} (#2) #3}
\def\pl#1#2#3{{\it Phys.\ Lett.} {\bf B#1} (#2) #3}
\def\physrev#1#2#3{{\it Phys.\ Rev.\ Lett.} {\bf #1} (#2) #3}
\def\prd#1#2#3{{\it Phys.\ Rev.} {\bf D#1} (#2) #3}
\def\ap#1#2#3{{\it Ann.\ Phys.} {\bf #1} (#2) #3}
\def\ppt#1#2#3{{\it Phys.\ Rep.} {\bf #1} (#2) #3}
\def\rmp#1#2#3{{\it Rev.\ Mod.\ Phys.} {\bf #1} (#2) #3}
\def\cmp#1#2#3{{\it Comm.\ Math.\ Phys.} {\bf #1} (#2) #3}
\def\mpla#1#2#3{{\it Mod.\ Phys.\ Lett.} {\bf #1} (#2) #3}
\def\jhep#1#2#3{{\it JHEP.} {\bf #1} (#2) #3}
\def\atmp#1#2#3{{\it Adv.\ Theor.\ Math.\ Phys.} {\bf #1} (#2) #3}
\def\jgp#1#2#3{{\it J.\ Geom.\ Phys.} {\bf #1} (#2) #3}
\def\int#1#2#3{{\it Int.J.Mod.Phys.}{\bf #1}:#2,#3}
\def\hepth#1{{\it hep-th/{#1}}}

\end{document}